%% file: acl_latex.tex
\pdfoutput=1
\documentclass[11pt]{article}

\usepackage[preprint]{acl}

\usepackage{times}
\usepackage{latexsym}

\usepackage[T1]{fontenc}
\usepackage[utf8]{inputenc}

\usepackage{microtype}

\usepackage{inconsolata}

\usepackage{graphicx}
\usepackage{CJKutf8}
\usepackage{enumitem}
\usepackage{multicol}
\usepackage{multirow}
\usepackage{xspace}
\usepackage{amssymb}
\usepackage{epigraph}
\usepackage{mathtools}
\usepackage{changepage}
\usepackage{listings}
\DeclareMathOperator*{\argmin}{arg\,min}

\usepackage{booktabs}

\title{Leveraging Uncertainty Estimation for Efficient LLM Routing}

\author{Tuo Zhang\thanks{These authors contributed equally to this work.} \\
  University of Southern California \\
  \texttt{tuozhang@usc.edu} \\\And
  Asal Mehradfar\footnotemark[1]\\
  University of Southern California \\
  \texttt{mehradfa@usc.edu} \\\AND
  Dimitrios Dimitriadis \\
  Amazon AI \\
  \texttt{dbdim@amazon.com} \\\And
  Salman Avestimehr \\
  University of Southern California \\
  \texttt{avestime@usc.edu} \\}

\begin{document}
\maketitle
\begin{abstract}
Deploying large language models (LLMs) in edge-cloud environments requires an efficient routing strategy to balance cost and response quality. Traditional approaches prioritize either human-preference data or accuracy metrics from benchmark datasets as routing criteria, but these methods suffer from rigidity and subjectivity. Moreover, existing routing frameworks primarily focus on accuracy and cost, neglecting response quality from a human preference perspective.
In this work, we propose the Confidence-Driven LLM Router, a novel framework that leverages uncertainty estimation to optimize routing decisions. 
To comprehensively assess routing performance, we evaluate both system cost efficiency and response quality. In particular, we introduce the novel use of LLM-as-a-Judge to simulate human rating preferences, providing the first systematic assessment of response quality across different routing strategies.
Extensive experiments on MT-Bench, GSM8K, and MMLU demonstrate that our approach outperforms state-of-the-art routing methods, achieving superior response quality while maintaining cost efficiency.
\end{abstract}

\input{sections/introduction}
\input{sections/system}

\input{sections/experiment}

\input{sections/conclusion}

\newpage
\input{sections/limitation}

\bibliography{custom}

\appendix
\input{sections/appendix}

\end{document}

%% file: sections/introduction.tex
\section{Introduction}
The deployment of AI models on edge devices is increasingly following a hybrid approach, where small language models (SLMs) run on-device (e.g., smartphones and IoT devices) while larger, more powerful models remain in the cloud~\cite{Gunter2024AppleIF}. This setup provides a balance between efficiency and performance, allowing low-latency responses for simple queries while reserving cloud-based LLMs for more complex tasks. However, determining when to offload queries to the cloud is a crucial challenge: calling the cloud unnecessarily increases cost and latency, whereas over-relying on local SLMs risks suboptimal response quality. An effective routing strategy is essential to dynamically balance cost and performance.

Traditional cascading routers, which sequentially query models until a satisfactory response is obtained~\cite{Chen2023FrugalGPTHT}, are inefficient for edge-cloud settings due to latency and redundant model calls. Recent predictive routing approaches aim to preemptively select the best model for a given query, with two leading solutions: TO-Router~\cite{Stripelis2024TensorOperaRA}, which trains router on accuracy-based benchmarks and RouteLLM~\cite{Ong2024RouteLLMLT}, which relies on human preference selection for router training.

\begin{figure}[t]
  \begin{center}
    \includegraphics[width=0.9\columnwidth]{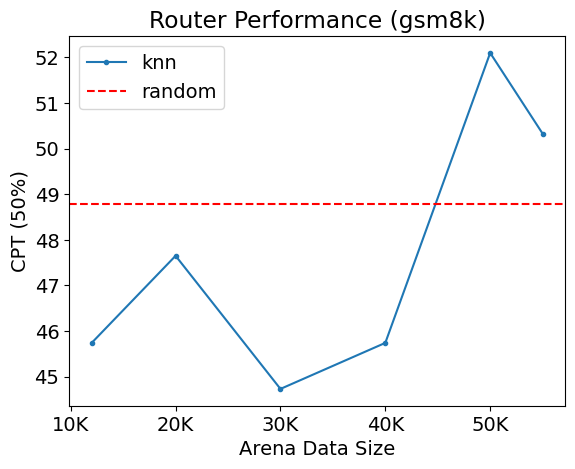}
  \end{center}
  \vspace{-5mm}
  \caption{Performance of human preference-based router with varying training sample sizes. Routing efficiency even becomes worse as the number of training samples increases, indicating that additional data does not necessarily improve performance.}
  \label{fig:scale_gsm8k}
  \vspace{-4mm}
\end{figure}

While human preference data and benchmark accuracy are commonly used as performance indicators in router training, our results reveal two major limitations that hinder data efficiency and system utilization. First, \textbf{\textit{human judgment is not always reliable.}} User ratings are subjective and inconsistent, often failing to provide an accurate ranking of model performance. Also, collecting human-preference data is resource-intensive, requiring manual evaluation of each sample on a case-by-case basis. These issues are particularly evident in the arena dataset~\cite{chiang2024chatbot}, where the distribution of preference data across models is sparse and uneven, complicating router training. Empirically, we demonstrate that arena data does not follow a scaling law to validate our argument. As shown in Figure~\ref{fig:scale_gsm8k}, increasing the dataset size does not necessarily improve routing performance. Instead, adding more data can introduce noise and inconsistencies, potentially degrading routing accuracy rather than enhancing it. Second, \textbf{\textit{accuracy is an incomplete indicator.}} Binary accuracy metrics do not capture response confidence, meaning two models may provide correct answers, yet one is significantly more precise and reliable.

\begin{figure}[t]
  \begin{center}
    \includegraphics[width=0.9\columnwidth]{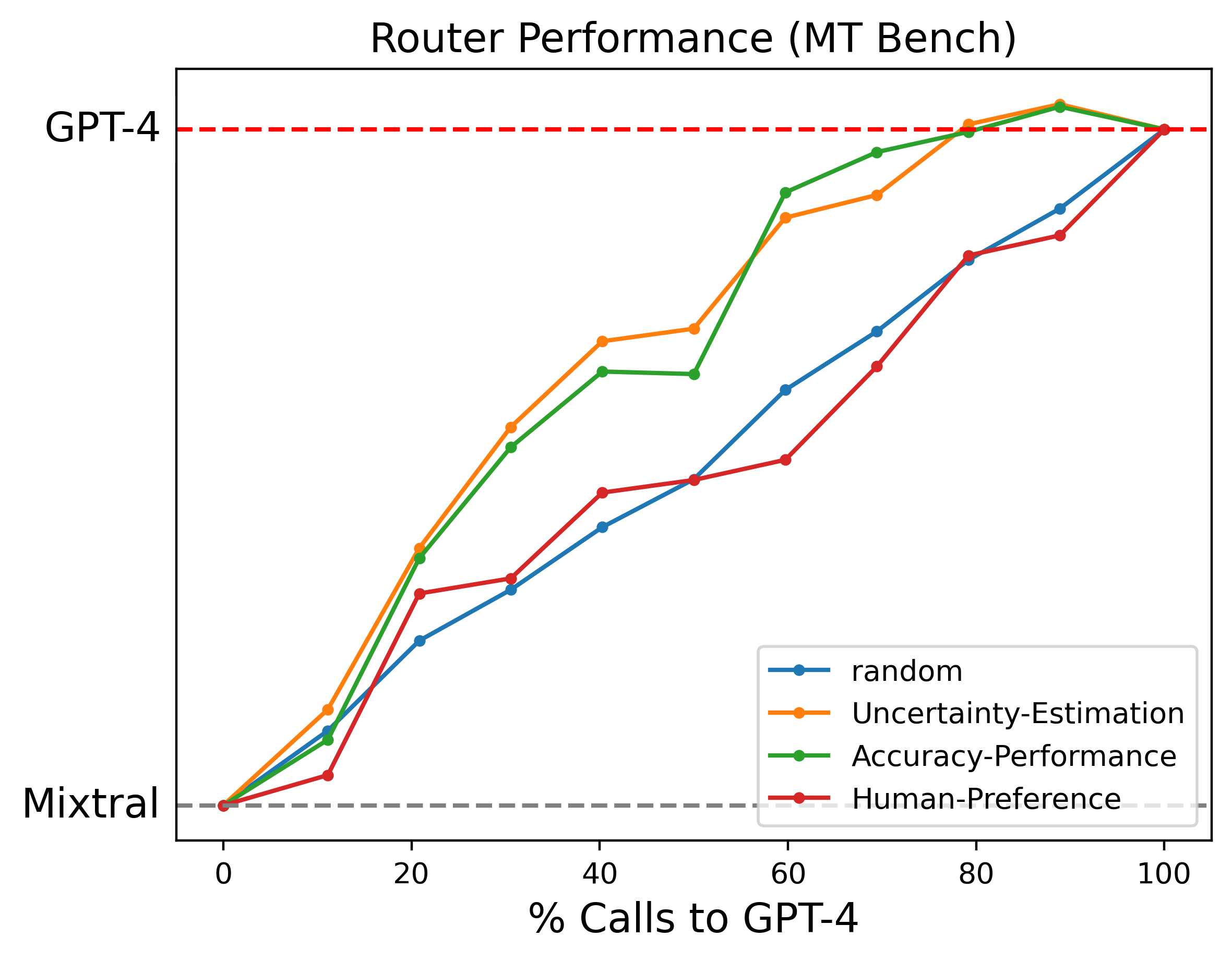}
  \end{center}
  \vspace{-5mm}
  \caption{Routing performance/cost trade-off between strong model (GPT-4) and weak model (Mixtral-8x7B). All routers shown, except the random router, use the same kNN-based model architecture.}
  \label{fig:mt_bench}
  \vspace{-4mm}
\end{figure}

To address the limitations of state-of-the-art methods, we propose \textbf{the Confidence-Driven LLM Router System}, which leverages Semantic Entropy (SE) as an uncertainty measure to guide routing decisions. Instead of relying on human preferences or accuracy-based thresholds, our system uses semantic entropy to measure model confidence. This enables the router to offload queries to cloud-based LLMs when higher certainty is needed, which keeps confident responses on-device to minimize cost. As the motivational example shown in Figure~\ref{fig:mt_bench}, by adopting uncertainty as a routing signal, our approach dynamically optimizes response quality and computational efficiency, making it better suited for real-world edge-cloud deployments.

%% file: sections/system.tex
\section{Method}
\subsection{Overview} \label{sec:system_overview}
Inspired by prior work~\cite{kuhn2023semantic, Bakman2024MARSMR}, we quantify uncertainty in natural language generation based on semantic content rather than token-level variations, using it as a performance indicator for training the router.

SE~\cite{kuhn2023semantic} captures uncertainty by clustering generated outputs with equivalent meanings and computing entropy over their aggregated probabilities. Unlike traditional entropy, which treats all token sequences distinctly, SE accounts for semantic equivalence, ensuring paraphrases contribute to the same uncertainty estimate. A lower SE score indicates higher confidence, while a higher score signals greater uncertainty.

Formally, we define the probability of a meaning cluster $c$ given an input prompt $\mathbf{x}$ as:

\begin{equation}
% \[
p(c | \mathbf{x}) = \sum_{\substack{s, \mathbf{x} \in c}} p(s | \mathbf{x}).
% \]
\label{equation:cluster}
\end{equation}
The semantic entropy of $\mathbf{x}$ is then computed as:
\begin{equation}
% \[
SE(\mathbf{x}) = -\frac{1}{|C|} \sum_{i=1}^{|C|} \log p(C_i | \mathbf{x}),
% \]
\label{equation:se}
\end{equation}
where $C$ represents the set of all clusters.

\subsection{System Design of the Confidence-Driven LLM Router}
The training and deployment of the Confidence-Driven LLM Router consist of three key phases:

\noindent \textbf{Phase 1: Router Data Preparation.} 
To create a training dataset that reflects real deployment scenarios, we select factual-related datasets, such as Natural QA and Trivia QA, to probe model confidence and knowledge capabilities. Additionally, domain-specific instruction datasets can be incorporated to tailor the router to specialized applications. 

Clustering generated outputs is a critical preprocessing step before computing SE. In this work, we adopt a bidirectional entailment mechanism following the previous work~\cite{kuhn2023semantic}. The first generated response initializes a cluster. For each subsequent response, a semantic entailment classifier, fine-tuned on DeBERTa-large model~\cite{He2020DeBERTaDB}, evaluates bidirectional entailment between the response and the current cluster representative. If both forward and backward entailments hold, the response is assigned to the existing cluster; otherwise, a new cluster is formed. This bidirectional criterion ensures that only semantically equivalent responses are grouped, allowing the number of clusters to be dynamically determined based on meaning variations among generated outputs.

\noindent \textbf{Phase 2: Constructing Preference Data from Semantic Entropy Scores.}
In the second phase, we create preference data by comparing the SE scores across models for each unique prompt. Although no two SE values are exactly the same, some prompts yield similar performance between models, which we denote as a “tie” case. To identify ties, we treat cases with close uncertainty levels as equivalent, where neither model is a clear winner.
To quantify the relative difference in uncertainty, we compute the normalized SE difference between the two models as:
\vspace{-1mm}
\begin{equation}
\delta_{\text{SE}}(\mathbf{x}) = \left| \frac{SE_{\text{strong}}(\mathbf{x}) - SE_{\text{weak}}(\mathbf{x})}{SE_{\text{strong}}(\mathbf{x})} \right|
\label{equation:normalized_se}
\end{equation}
$SE_{\text{strong}}(\mathbf{x})$ represents the high-cost model, and $SE_{\text{weak}}(\mathbf{x})$ represents the low-cost model. Using this metric, we determine the preferred model as:
\begin{equation} 
\text{Winner} = 
\begin{cases} 
\arg\min\limits_{M} SE(M, \mathbf{x}) & \text{if } \delta_{\text{SE}}(\mathbf{x}) > \tau, \\
\text{Tie} & \text{otherwise}.
\end{cases} 
\label{equation:threshold} 
\end{equation}
The predefined threshold $\tau$ controls sensitivity to uncertainty differences. If $\delta_{\text{SE}}(\mathbf{x})$ exceeds the predefined $\tau$, the uncertainties are considered sufficiently distinct, and the model with the lower semantic entropy is designated as the preferred model. Otherwise, we consider both models equal in performance for the given prompt.

A higher $\tau$ enforces stricter distinctions, aligning the preference data more closely with traditional accuracy-based evaluations. Conversely, a lower $\tau$ increases sensitivity to subtle linguistic variations in model responses. By tuning $\tau$, we balance robustness with linguistic granularity.

\noindent \textbf{Phase 3: Training the Confidence-Driven Router.}
After generating the SE-based preference data in Phase 2, we format each training sample as follows: \{id, model a, model b, prompt, response a, response b, winner model a, winner model b, winner tie\}. In the last three columns, we use binary values (0 or 1) to represent the routing outcomes. For instance, if model\_a is the preferred model, then \texttt{winner\_model\_a} is set to 1, while \texttt{winner\_model\_b} and \texttt{winner\_tie} are set to 0. The dataset used in this study is now publicly available on Huggingface~\footnote{\url{https://huggingface.co/datasets/AsalMehradfar/uncertainty_0.1}}. Once the dataset is prepared, we transform the instruction records into vectorized representations using the pre-trained embedding model, which serves as inputs for training the router classifiers.

% ~\footnote{The training dataset would be available when the paper is public.}. 

%% file: sections/experiment.tex
\section{Evaluation}
\subsection{Experimental Methodology}
\subsubsection{Tasks, Datasets, and Models}
We use GPT-4 as the strong model and Mixtral-8x7B as the weaker model.
The Confidence-Driven Router is trained with a combination of Natural QA~\cite{Kwiatkowski, lee-etal-2019-latent}, Trivia QA~\cite{2017arXivtriviaqa}, PopQA~\cite{mallen2023llm_memorization}, and MAWPS~\cite{KoncelKedziorski2016MAWPSAM} datasets to ensure a knowledge space. We randomly selected 3,610 samples from each QA dataset and 1,418 samples from the MAWPS dataset, resulting in 12,247 samples, matching the quantity as Chatbot Arena dataset~\cite{chiang2024chatbot} for RouteLLM training. We select MMLU~\cite{Hendrycks2020MeasuringMM}, MT-Bench~\cite{Zheng2023JudgingLW}, and GSM8K datasets~\cite{Cobbe2021TrainingVT} to comprehensively evaluate the routing systems. Following the previous works~\cite{Ong2024RouteLLMLT, Stripelis2024TensorOperaRA}, we select four different predictive routing models for evaluation: Similarity-Weighted (SW) Ranking, Matrix Factorization (MF), Multilayer Perceptron (MLP), and k-Nearest Neighbors (kNN).
More details related to the datasets and routing models are listed in the Appendix.

\begin{table*}[t]
\caption{System Performance Comparison of Routing Systems on test datasets. A low CPT value indicates a cost-effective routing strategy. \textbf{Bold} highlight the best performance, and \underline{underlined} denote the second-best.}

    % \vspace{1mm}
    \footnotesize
    \begin{tabular*}{1.0\linewidth}{cccccccc}
        \toprule
        & &
        \multicolumn{2}{c}{\textbf{MT-Bench}} & 
        \multicolumn{2}{c}{\textbf{GSM8K}} &
        \multicolumn{2}{c}{\textbf{MMLU}} \\ 
        \midrule 

        \textbf{Routing} & 
        \textbf{Method} & 
        \textbf{CPT(50\%)} & 
        \textbf{CPT(80\%)} & 
        \textbf{CPT(50\%)} & 
        \textbf{CPT(80\%)} &  
        \textbf{CPT(50\%)} & 
        \textbf{CPT(80\%)} \\
        \midrule 

        &
        Random &
        51.29 &
        78.55 &
        48.79 &
        80.16 &
        50.04 &
        79.32 \\
        \midrule

        \multirow{2}{*}{TO-Router} &
        kNN &
        59.72 &
        90.39 &
        47.93 &
        79.03 &
        43.73 &
        77.74 \\

        &
        MLP &
        49.15 &
        86.67 &
        51.03 &
        77.77 &
        44.01 &
        77.43 \\
        \midrule

        \multirow{2}{*}{RouteLLM} &
        SW &
        56.08 &
        78.37 &
        46.03 &
        79.58 &
        47.41 &
        \underline{74.23}\\

        &
        MF &
        55.59 &
        84.12 &
        49.07 &
        80.09 &
        58.55 &
        83.68 \\
        \midrule

        \multirow{4}{*}{
        \shortstack{Confidence-Driven \\ LLM Router}}
        &
        SW &
        \textbf{27.31} &
        \textbf{55.61} &
        48.03 &
        80.41 &
        \textbf{37.96} &
        \textbf{73.85} \\

        &
        MF &
        42.94 &
        \underline{63.53} &
        \textbf{41.89} &
        \textbf{75.34} &
        50.06 &
        78.38\\

        &
        kNN &
        60.84 &
        81.50 &
        \underline{44.08} &
        \underline{76.32} &
        \underline{42.70} &
        75.28\\

        &
        MLP &
        \underline{35.54}&
        74.92&
        50.04&
        79.79&
        57.07&
        83.27\\
        \bottomrule
    \end{tabular*}
\vspace{-1mm}
\label{table:baseline_results}
\end{table*}

\subsubsection{Evaluation Criteria}
We evaluate performance based on two key criteria: system cost and response quality.

To evaluate system cost, we adopt the Call-Performance Threshold (CPT) metric from prior work~\cite{Ong2024RouteLLMLT}. CPT(x\%) represents the minimum fraction of queries that must be routed to the stronger model to achieve an x\% improvement over the baseline accuracy of the weaker model. For example, CPT(50\%) specifies the proportion of calls required to the strong model to attain a 50\% improvement over the weak model’s baseline accuracy. A lower CPT value indicates a more cost-effective routing strategy, achieving performance gains with fewer calls to the stronger model.

To evaluate response quality, we use LLM-as-a-Judge to simulate human ratings. We employ an independent LLM (i.e., GPT-o1) to choose the most preferable responses from the routed model alongside ground-truth answers. The judge is instructed to select based on correctness and precision of reasoning, as the detailed system prompt listed in the Appendix. We design the score as $\text{Score}(i) = \left(\frac{S_i}{T}\right) \times 100$, where \( S_i \) be the number of times router \( i \) is selected, and \( T \) be the total number of queries. Unlike traditional accuracy-based evaluations using golden labels, this approach simulates human judgment by considering not only the correctness but also the interpretability and coherence of model outputs, better aligning with human preference selection rather than relying solely on objective correctness.

\begin{table}[t]
% \scriptsize
% \footnotesize
\begin{center}
\caption{
Response quality comparison of routing systems on the GSM8K dataset. Higher LLM-judge scores reflect better response quality.
}
\vspace{-2mm}
\resizebox{1.0\columnwidth}{!}{
\begin{tabular}{cccc} 
\hline
\toprule
& TO-Router & RouteLLM & Confidence Router \\ %\hline
\midrule
CPT(50\%) & 78.88 & 79.72 & \textbf{79.95} \\
\midrule
CPT(80\%) & 85.97 & 88.88 & \textbf{89.21} \\
\bottomrule
\label{tab:se_results}
\end{tabular}
}
\end{center}
\vspace{-7mm}
\end{table}

\subsubsection{Baseline Selection}
We select RouteLLM~\cite{Ong2024RouteLLMLT} and TO-Router~\cite{Stripelis2024TensorOperaRA}, two state-of-the-art predictive routing systems. Following their original configurations, we evaluate TO-Router using kNN and MLP architectures, while RouteLLM is assessed using SW ranking and MF models. We also include a random router without any training as a baseline for comparison.

\subsection{Evaluation on System Costs}
We first evaluate the system costs of various routing strategies on the test datasets, as summarized in Table~\ref{table:baseline_results}. The Confidence-Driven LLM Router consistently achieves strong performance in CPT(50\%) and CPT(80\%) across datasets.
To provide a clearer comparison, we report the actual OpenAI API cost (in USD) for each routing system on the MT-Bench dataset under CPT(80\%): Random router costs \$4.06; TO-Router costs \$3.88; RouteLLM costs \$4.04; and our proposed method costs lowest with \$3.74. 
These results demonstrate that leveraging uncertainty as a routing indicator provides notable advantages in achieving target performance improvements while balancing computational costs.

\subsection{Evaluation on Response Quality}

To better understand the advantage of our proposed method, we evaluate the response quality of each routing system under the same accuracy with GSM8K dataset. We select the best routing models under each routing system in Table~\ref{table:baseline_results}. As summarized in Table~\ref{tab:se_results}, the proposed method achieves the highest LLM-as-a-Judge rating, indicating that its responses are the most human-preferable among all baselines. Our uncertainty-aware training strategy optimizes routing decisions based on a direct measure of model confidence. By minimizing uncertainty in routing, our approach ensures that queries are directed to models that can generate the most confident and reliable responses, leading to outputs that better align with human preferences.

%% file: sections/conclusion.tex
\section{Conclusion}
In this paper, we introduced the Confidence-Driven LLM Router, a novel framework that leverages uncertainty estimation to optimize LLM deployment in edge-cloud environments. 
By using semantic entropy as a performance indicator, our approach addresses the limitations of existing methods, such as the subjectivity of human preference data and the rigidity of accuracy-based metrics. 
Extensive experiments on benchmark datasets demonstrate that our method outperforms state-of-the-art routing systems, achieving a better trade-off between response quality and efficiency. Future work will explore multi-modal query integration and further latency reduction in distributed systems.

%% file: sections/limitation.tex
\section{Limitations}
We have two major limitations which we aim to address in future works. 

First, our evaluation is limited to text-based queries, and we do not extend our analysis to multi-modal routing scenarios. In real-world applications, queries may include image-text pairs or other modalities, especially in Vision-Language Models (VLMs). Future work should investigate how uncertainty estimation-based routing generalizes to multi-modal inputs and whether SE remains an effective performance indicator in VLM settings.

Second, we do not analyze the computational overhead of different routing architectures. Our study primarily evaluates routing effectiveness, but in practice, the choice of router architecture (e.g., SW, MF, kNN, and MLP) can significantly impact system efficiency. Future work should explore the trade-offs between neural network-based routers and statistical methods, assessing their cost, scalability, and real-time deployment feasibility in edge-cloud environments.

\section{Ethical Statement}
This work does not raise any ethical concerns.

%% file: sections/appendix.tex
\newpage
\onecolumn
\section{Appendix}
\subsection{Details about Routing Architectures}
We select four different predictive routing methods in our evaluation. To match the hardware constraints on edge computing, we purposely select the lightweight routing architectures in our experiments. Now, we describe our approach for learning the win prediction model \( P(\text{win}_{\mathcal{M}_{\text{strong}}} \mid q) \). We represent a sample from our dataset \( \mathcal{D} \) as \( (q, M_w, M_l) \), where \( M_w \) and \( M_l \) denote the winning and losing models, respectively.

\textbf{Similarity-Weighted (SW) Ranking.} 
Same as RouteLLM \cite{Ong2024RouteLLMLT}, we adopt a Bradley-Terry (BT) model \cite{Bradley1952RankAO} for this routing task. Given an input query \( q \), we compute a weight \( \omega_i \) for each query \( q_i \) in the training set based on its similarity to \( q \), as follows:
\begin{equation}
\omega_i = \gamma^{1 + S(q, q_i)},
\end{equation}
where \( \gamma \) is a scaling factor which is 10 in our case, and \( S(q, q_i) \) represents the similarity between queries \( q \) and \( q_i \), defined as:
\begin{equation}
S(q, q_i) = \frac{\epsilon \cdot \epsilon_i}{\|\epsilon\| \|\epsilon_i\| \cdot \max_{1 \leq s \leq |D|} \left( \frac{\epsilon_i \cdot \epsilon_s}{\|\epsilon_i\| \|\epsilon_s\|} \right)} ,
\end{equation}
with \( \epsilon \) denoting a query embedding. The BT coefficients \( \xi \) are then learned by solving:
\begin{equation}
\arg\min_{\xi} \sum_{i=1}^{|D|} \omega_i \cdot \ell \left( l_i, \frac{1}{1 + e^{\xi_{w_i} - \xi_{l_i}}} \right),
\end{equation}
where \( \ell \) is the binary cross-entropy loss.

The learned BT coefficients allow us to estimate the win probability given query \( q \) as:
\begin{equation}
P(\text{win}_{M_w} \mid q) = \frac{1}{1 + e^{\xi_w - \xi_l}}.
\end{equation}
This routing model does not require additional training, and the optimization is performed at inference time.

\textbf{Matrix Factorization.} Inspired by the RouteLLM approach \cite{Ong2024RouteLLMLT, 5197422}, we leverage matrix factorization as one of our routing models. The objective is to uncover a latent scoring function \( s: \mathcal{M} \times \mathcal{Q} \rightarrow \mathbb{R} \) that assesses the quality of the model \( M_w \)'s response to a given query \( q \). Specifically, if model \( M_w \) performs better than \( M_l \) on query \( q \), then \( s(M_w, q) > s(M_l, q) \). We encode this preference by modeling the win probability using a Bradley-Terry (BT) relationship \cite{Bradley1952RankAO}:
\begin{equation}
P(\text{win}_{M_w} \mid q) = \sigma(s(M_w, q) - s(M_l, q)),
\end{equation}
where \( \sigma \) is the sigmoid function, and \( s \) is a bilinear function of the model and query embeddings. This approach effectively performs matrix factorization over the score matrix on the set \( \mathcal{Q} \times \mathcal{M} \).

\textbf{Multilayer Perceptron (MLP).} For the MLP routing, we utilize a 2-layer multilayer perceptron (MLP) architecture. The output \( y_k \) for the MLP-Router is given by:
\begin{equation}
P(\text{win}_{M_w} \mid q) = \varphi \left( \sum_{j=1}^{m} w_{jk}^{(2)} \text{ReLU} \left( \sum_{i=1}^{n} w_{ij}^{(1)} \epsilon_i + b_j^{(1)} \right) + b_k^{(2)} \right),
\end{equation}
where \( \varphi \) represents the softmax activation function in the output layer and \( \epsilon \) denoting a query embedding.

\textbf{k-Nearest Neighbors (kNN).}
The k-Nearest Neighbors router represents all training queries \( q_i \) with an embedding \( \epsilon_i \). For each test query \( q \), with embedding \( \epsilon\), we identify the closest training query \( q' \) by finding the query in the training set with the highest cosine similarity to \( \epsilon \):
\begin{equation}
i = \argmin_{1 \leq i \leq |D|} \left( \frac{\epsilon_i \cdot \epsilon}{\|\epsilon_i\| \|\epsilon\|} \right).
\end{equation}
\begin{equation}
q' = q_i
\end{equation}
After identifying the nearest query \( q' \), the router decides on the winner model based on the performance of the winner model associated with \( q' \). 
This method leverages the similarity between the test query and training queries to select the most suitable expert dynamically.

\subsection{Details about Datasets}
To train the RouteLLM router, we randomly sample 12,247 data points from the Chatbot Arena dataset~\cite{chiang2024chatbot}. In contrast, both the Confidence-Driven Router and TO-Router are trained using a combination of Natural QA~\cite{Kwiatkowski, lee-etal-2019-latent}, Trivia QA~\cite{2017arXivtriviaqa}, PopQA~\cite{mallen2023llm_memorization}, and MAWPS~\cite{KoncelKedziorski2016MAWPSAM} datasets to ensure a comparable data volume. Specifically, we randomly selected 3,610 samples from each QA dataset and 1,418 samples from the MAWPS dataset, resulting in 12,247 samples, matching the quantity used for RouteLLM.

To comprehensively evaluate the routing systems, we select a diverse set of benchmark datasets: the MMLU~\cite{Hendrycks2020MeasuringMM} dataset, consisting of 14,042 questions across 57 subjects; the MT-Bench dataset~\cite{Zheng2023JudgingLW}, which includes 160 open-ended questions assessed using the LLM-as-a-judge approach; and the GSM8K dataset~\cite{Cobbe2021TrainingVT}, containing over 1,000 grade-school math problems. These datasets provide a broad evaluation across varied question types and subject domains. For all the data listed above, we properly use them under the propose of research by following their license. 

\subsection{Models and Test Environment}
We implemented the experiments using PyTorch~\cite{pytorch}, and conducted our experiments on two NVIDIA A100 GPUs. For GPT-4 model, we use \texttt{gpt-4-0613} API. For GPT-o1 model, we use \texttt{o1-2024-12-17} API.

\subsection{System Prompt Design for LLM-as-a-Judge}
To evaluate response quality, we use LLM-as-a-Judge to simulate human ratings. We employ an independent LLM (i.e., GPT-o1) to choose the most preferable responses from the routed model alongside ground-truth answers. The judge is instructed to select based on correctness and precision of reasoning. We designed and implemented the following system prompt:

\begin{adjustwidth}{1cm}{1cm}
\begin{lstlisting}[breakatwhitespace=true, showstringspaces=false, basicstyle=\ttfamily, columns=fullflexible, breaklines=true]
You are an evaluator for math problem solutions. You will receive:
1. A question.
2. A ground truth answer.
3. Three LLM-generated responses.
Your task is to select which response(s) is/are best, based on whether the answer is correct and the reasoning is precise.
Follow these rules: 
* DO NOT provide any explanation or reasoning in your answer-only state which LLM(s) you judge as having the best response.
* If more than one response is equally best, name each of them.
Question: {}
Ground Truth Answer:{}
LLM 1 Response: {}
LLM 2 Response: {}
LLM 3 Response: {}
Your output must ONLY indicate the selected LLM(s). For example, 'LLM 1' or 'LLM 1 and LLM 3'.
\end{lstlisting}
\end{adjustwidth}